\documentclass[10pt,twocolumn,showpacs,amsmath,amssymb,osajnl,floatfix]{revtex4-1}
\usepackage{epsfig}
\usepackage{amsmath}
\usepackage{amssymb}
\usepackage{bm}
\usepackage{graphicx}
\usepackage{color}
\usepackage{graphicx,rotating}
\usepackage{txfonts}
\usepackage{microtype}
\usepackage{ulem}

\begin{document}
\author{Jian-Xing Li}\email{Jian-Xing.Li@mpi-hd.mpg.de}
\author{Yue-Yue Chen}
\author{Karen Z. Hatsagortsyan}\email{k.hatsagortsyan@mpi-k.de}
\author{Christoph H. Keitel}
\affiliation{Max-Planck-Institut f\"{u}r Kernphysik, Saupfercheckweg 1,
69117 Heidelberg, Germany}

\bibliographystyle{apsrev4-1}

\title{Angle-resolved stochastic photon emission  in the quantum radiation-dominated regime}

\date{\today}

\begin{abstract}
 
Signatures of stochastic effects in the radiation of a relativistic electron beam interacting with a counterpropagating superstrong short focused laser pulse are investigated in a quantum regime when the electron's radiation dominates its dynamics. We consider the electron-laser interaction at near-reflection conditions when pronounced high-energy gamma-ray bursts arise in the backward-emission direction with respect to the initial motion of the electrons. The quantum stochastic nature of  the gamma-photon emission is exhibited in the angular distributions of the radiation and explained in an intuitive picture.  Although, the visibility of the stochasticity signatures depends on the laser and electron beam parameters, the signatures are of a qualitative nature and robust. The stochasticity, a fundamental quantum property of photon emission,  should thus be measurable rather straightforwardly with laser technology available in near future.


\end{abstract}

\maketitle

The next generation petawatt laser systems \cite{ELI,HiPER,XCELS} will open a door not only to novel regimes of laser-matter interaction \cite{Sarri_2014, Sarri_2015}, but also to new perspectives for the investigation of fundamental problems \cite{Marklund_2006,Mourou_2006,Esarey_2009,RMP_2012,Mourou_2014}. In ultrastrong laser fields the quantum properties of electron radiation, the discrete and probabilistic character of photon emission, can be conspicuous.  While the discreteness of the radiation photon energy is known to be observed straightforwardly, e.g., in the Compton scattering as a shift of the emission frequency \cite{Goldman_1964,Brown_1964,Landau_4}, 
the signatures of the stochastic character of photon emission are more subtle and elaborate for observation. The latter has impact on the radiation back-action to the electron dynamics and should be more apparent in the so-called radiation dominated regime (RDR) of interaction \cite{Koga_2005,DiPiazza_2009,RMP_2012}, when multiple emission of photons by an electron becomes probable. One of conceptual consequences of 
stochasticity effects (SE) in photon emissions is the broadening of the energy spread of an electron beam in a plane laser field \cite{Neitz_2013,Neitz_2014}, while similar effect can cause electron stochastic heating  in a standing laser field \cite{Bashinov_2015}. However, in an experiment in a focused laser beam, competing  effects may arise, e.g., an additional energy spreading of the electron beam  due to the difference of radiative losses of electrons in the electron-beam cross section. Another SE signature is  the so-called electron straggling effect during radiation in strong fields \cite{Shen_1972,Duclous_2011,Blackburn_2014,Harvey_2016b},  when the electrons propagate a long distance without radiation due to SE, resulting in the increase of the yield of high-energy photons.  However, we will show below that the straggling effect is rather weak in the considered regime.
Different signatures of quantum radiation reaction have been discussed \cite{Sokolov_2009,Sokolov_2010b,Sokolov_2010,DiPiazza_2010,Bulanov_2013,Jian-Xing_2014,Green_2014,Jian-Xing_2015,Wang_2015,Yoffe_2015,Vranic_2016,Harvey_2016,Seipt_2016,Dinu_2016} which include all quantum effects, such as photon recoil, stochasticity, and interferences \cite{Dinu_2016}. However, they do not isolate  information on the specific role of stochasticity. Unfortunately, so far unequivocal SE have not been observed in an experiment.

The quantum effects in strong laser fields are determined by the invariant parameter $\chi\equiv |e|\sqrt{(F_{\mu\nu}p^{\nu})^2}/m^3$ \cite{Reiss_1962, Ritus_1985}, where $F_{\mu\nu}$ is the field tensor, $p^{\nu}=(\varepsilon,\textbf{p})$  the incoming electron 4-momentum, and $e$ and $m$ are the electron charge and mass, respectively (Planck units $\hbar=c=1$ are used throughout). The RDR, when the radiation losses during a laser period are comparable with the electron initial energy,
is characterized by the parameter $R \equiv\alpha \xi \chi \gtrsim 1$ \cite{RMP_2012}, where $\alpha$ is the fine structure constant,  $\xi\equiv |e|E_0/(m\omega_0)$ the invariant laser field parameter, while $E_0$ and $\omega_0$ are the laser field amplitude and frequency, respectively.   
In the quantum RDR the stochastic nature of photon emission is a fundamental quantum property and has  to be taken into account during the high-energy photon radiation \cite{Nerush_2011,Elkina_2011,Ridgers_2014,Green_2015}, that significantly affects the electron dynamics and the high-energy photon emission.

In this letter, we investigate  signatures of the stochastic nature of photon emission in the nonlinear Compton scattering in the quantum RDR during the interaction of a superstrong short focused laser pulse with a counterpropagating relativistic electron beam. We consider the interaction in the electron near-reflection regime \cite{DiPiazza_2009}: due to the combined effects of the laser focusing and radiation reaction, the front electrons of the electron beam are reflected and emit ultrashort gamma-rays in the near-backward direction with respect to the initial electron motion \cite{Jian-Xing_2015}. In the considered case with $\chi \lesssim 1$, the straggling effect is rather weak, however, the  electron near-reflection regime offers a possibility to observe the SE in the angular distribution of the radiation. In fact, the photon emissions in different laser cycles are essentially modified due to the SE in the RDR, and the latter is mapped into the broad backward-emission angles when the electron is at the near-reflection condition. We calculate angle-resolved radiation intensity and photon numbers and show that, due to the stochastic nature of photon emission, the radiation's angular distribution (RAD) has a single prominent peak in the backward direction, which is broad and easily observable in an experiment. In contrast, when the SE are ignored, the backward radiation yields an angular distribution with several peaks. Furthermore, we investigate the influences of the  laser and electron-beam parameters (the laser focal radius, the laser pulse duration and the electron initial energy) on the visibility of those SE signatures,  to optimize parameters for a future experimental setup.

The applied parameter domain is defined as follows. The considered quantum RDR requires the invariant parameters $\chi\equiv \gamma(\omega_0/m)\xi (1-\beta \cos\theta)\approx10^{-6}\gamma\xi\lesssim 1$ and $R \equiv\alpha \xi \chi \gtrsim 1$, while the electron near-reflection regime does $\gamma\sim \xi/2$, where $\gamma$ is the Lorenz factor of the electron, and $\beta$ the electron velocity scaled by the light speed in vacuum. The two conditions above demand $\gamma\sim \xi\sim 10^3$, i.e., an electron beam of GeV energies and laser intensities of $10^{23}$-$10^{24}$ W/cm$^2$ anticipated in next generation facilities  \cite{ELI,HiPER,XCELS}.

The calculation of the radiation is based on Monte-Carlo simulations employing QED theory for the electron radiation and classical equations of motion for the propagation of electrons between photon emissions \cite{Elkina_2011,Ridgers_2014,Green_2015}. 
In superstrong laser fields $\xi\gg 1$, the coherence length of the photon emission is much smaller than the laser wavelength  and the typical size of the electron trajectory \cite{Ritus_1985,Khokonov_2010}. Then, the photon emission probability is determined by the local electron trajectory, consequently, by the local value of the parameter $\chi$ \cite{Baier_b_1994}. The radiation in the quantum regime ignoring the SE  are calculated by employing the semi-classical method \cite{Sokolov_2009,Sokolov_2010,Sokolov_2010b}, when the radiation is emitted continuously along the electron trajectory and modifies  accordingly the classical equations of motion \cite{Suppl_material}.

We employ a linearly polarized tightly focused  laser pulse with a Gaussian temporal profile \cite{Suppl_material}, which propagates along $+z$-direction and  polarizes in $x$-direction.
 The spatial distribution of the electromagnetic fields takes into account up to the $\epsilon^3$-order of the nonparaxial solution \cite{Salamin_2002, Salamin_2007}, where $\epsilon=w_0/z_r$, 
$w_0$ is the laser focal radius, $z_r = k_0w_0^2/2$  the Rayleigh length with the laser wave vector $k_0=2\pi/\lambda_0$, and $\lambda_0$  the laser wavelength. 

\begin{figure}[t]
\includegraphics[width=8cm]{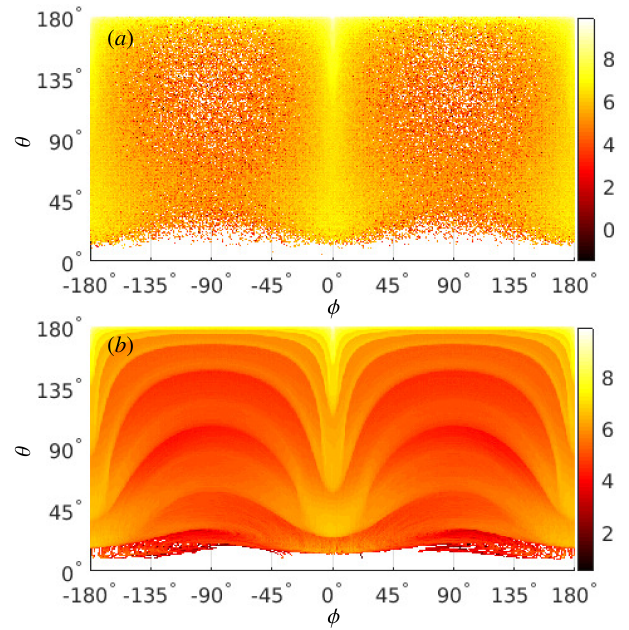}\\
\caption{(Color online) Angle-resolved radiation energy $\varepsilon_R$ in units of the electron rest energy $m$ vs the emission polar angle $\theta$ and the azimuthal angle $\phi$: (a) including and (b) excluding  SE, in a 6-cycle focused laser pulse. Color coded is log$_{10}$[d$\varepsilon_R/$d$\Omega$] rad$^{-2}$, with the emission solid angle $\Omega$. The laser and electron beam counterpropagate, with the  initial propagation polar angles $\theta_L^{(i)}=0^{\circ}$ and $\theta_e^{(i)}=180^{\circ}$, respectively. All other parameters are given in the text.  
 }
\label{fig1}
\end{figure}

A typical angular distribution of radiation  which carries the signature of the stochastic nature of photon emission, is illustrated in Fig.~\ref{fig1}; $\theta$ is  the emission polar angle   with respect to the laser propagation direction, and $\phi$  the azimuthal angle; $\phi=0^{\circ}$ and $\pm 180^{\circ}$ correspond to the positive and negative directions of the laser polarization, respectively. The  peak intensity of the 6-cycle (FWHM) laser pulse is $I\approx 4.9\times 10^{23}$W/cm$^2$ ($\xi=600$), $\lambda_0 = 1$ $\mu$m, and $w_0$ = 2 $\mu$m. The electron beam, with radius $w_e=\lambda_0 $, length $L_e=6 \lambda_0$, and density $n_e\approx 10^{15}$ cm$^{-3}$, initially counterpropagates with the laser pulse,  i.e. $\theta_e^{(i)}=180^{\circ}$. The initial mean kinetic energy of the electron beam is $\varepsilon_0 =180$ MeV ($\gamma_0\approx353$, the maximum value of $\chi$ during interaction $\chi_{max}\lesssim 1$), and the energy and angular spread  are $\Delta \varepsilon/\varepsilon_0=\Delta \theta=10^{-3}$. The electron-beam parameters are typical for  current laser-plasma acceleration setups \cite{Marklund_2006,Mourou_2006,Esarey_2009}.

Figures~\ref{fig1}(a) and \ref{fig1}(b) demonstrate RAD including and excluding  SE, respectively.  The radiation is  most significant along the strongest field component $E_x$ of the linearly polarized laser field. However, in a tightly focused laser beam, other components of the electric field, $E_y$ and $E_z$,  are not negligible and play a significant role in the electron dynamics and the  photon emission. Consequently, the electrons radiate continuously over the azimuthal angle $\phi$, with  a Gaussian radiation distribution with respect to $\phi$ corresponding to the laser transverse profile. Moreover, the radiation sweeps from the polar angle $\theta=180^{\circ}=\theta_e^{(i)}$ down to $\theta\approx 11^{\circ}$. The electrons initially counterpropagate  with the laser pulse,  emit forwards,  and create a high-intensity spectral region around $\theta\approx 180^{\circ}$. Due to radiative losses, the electron energy decreases, facilitating the reflection condition $\gamma\approx \xi/2$ in the strong-field region and inducing  electron reflection, during which  the radiation flash sweeps down to $\theta \approx 11^{\circ}$. 
The emission angle after the reflection with respect to the electron average motion $\delta\theta  \sim \xi/\gamma$ is determined by the values of $\xi$ and $\gamma$ \textit{in situ} after the reflection. In the region of $\theta\approx 0^{\circ}$, the radiation intensity is vanishing, because of the very low $\chi$ value for the co-propagating electron: the radiation energy $\varepsilon_R\varpropto\chi\varpropto\gamma\xi$(1-$\beta$cos$\theta)\approx 0$ at $\theta\approx 0^{\circ}$.

The  radiation distributions with respect to the polar angle $\theta$ with and without SE are essentially different  in Fig.~\ref{fig1}. 
While with SE the RAD is smoothly peaking at $\theta=180^{\circ}$ and at small angles, without SE it shows a band structure corresponding to the radiation emerging from different laser cycles. A similar behaviour is seen in RAD for photon numbers \cite{Suppl_material}.

\begin{figure}[t]
\includegraphics[width=8cm]{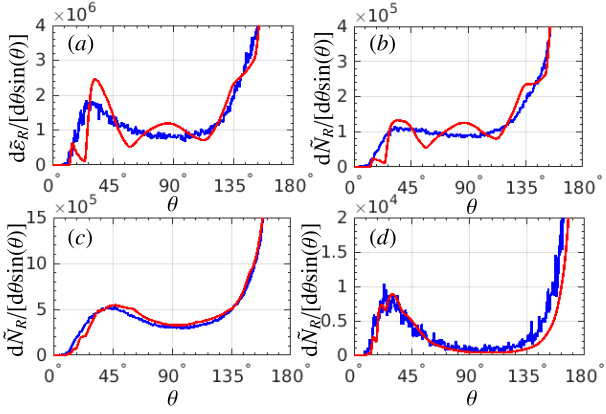}\\
\caption{(Color online) RAD: (a) radiation energy d$\tilde{\varepsilon}_R$/[d$\theta$sin($\theta$)] (rad$^{-1}$), and (b) photon number d$\tilde{N}_R$/[d$\theta$sin($\theta$)] (rad$^{-1}$),  emitted in the region of $-15^{\circ}\leq \phi\leq +15^{\circ}$ of the azimuthal angle, see Fig.~\ref{fig1}. The emitted photon number d$\tilde{N}_R$/[d$\theta$sin($\theta$)] in the whole $2\pi$-region of the azimuthal angle: (c) all photon energies, and (d)  the photon energies above 50 MeV. Blue curves are with SE, and red curves are without SE. The employed laser and electron beam parameters are the same as in Fig.~\ref{fig1}. }
\label{fig2}
\end{figure}

A quantitative comparison  between RADs including and excluding  SE is represented in Fig.~\ref{fig2}. We focus on the strongest radiation domain along the polarization plane in the region of $-15^{\circ}\leq \phi\leq +15^{\circ}$, analysing the radiation energy   d$\tilde{\varepsilon}_R$/[d$\theta$sin($\theta$)]= $\int_{-15^{\circ}}^{+15^{\circ}}$ d$\phi$  d$\varepsilon_R$/d$\Omega$ and the photon number d$\tilde{N}_R$/[d$\theta$sin($\theta$)]= $\int_{-15^{\circ}}^{+15^{\circ}}$ d$\phi$  d$N_R$/d$\Omega$ in this domain. The stochastic nature of photon emission is clearly discernible in RAD: a single broad high-intensity gamma-photon peak is formed in the near-reflection direction when SE is included, while in the case without SE multiple radiation peaks emerge corresponding to the emission from different laser cycles. 
 
  \begin{figure}[t]
\includegraphics[width=8cm]{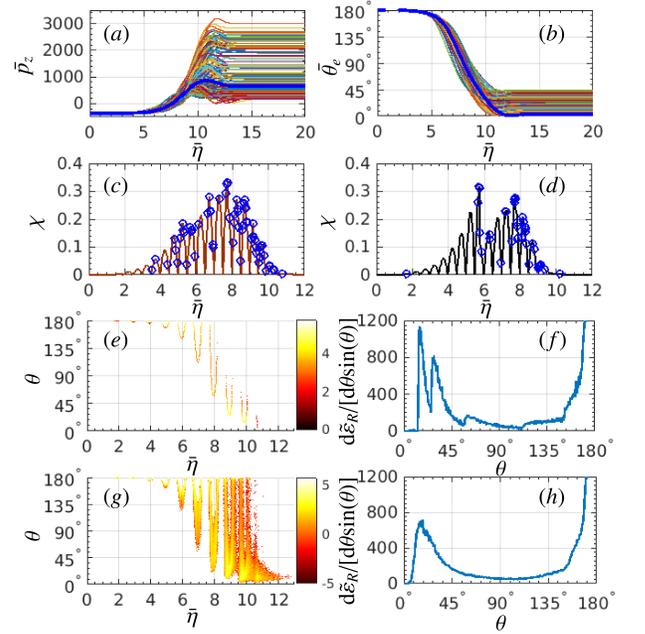}\\
\caption{(Color online) (a) The mean longitudinal momentum $\bar{p_z}$ and (b) the mean motion direction (photon-emission direction) $\bar{\theta_e}$ (polar angle)   of a sample electron. The thick line is without SE, and  thin lines are with SE. (c) and (d): The variation of the parameter $\chi$ with respect to $\bar{\eta}$ of a sample electron with SE. The blue marks indicate the photon emissions.
The single electron radiation, integrated over the azimuthal angle of $-15^{\circ}\leq \phi\leq +15^{\circ}$, without SE: (e)  radiation intensity vs emission phase $\bar{\eta}$,   log$_{10}$[d$^2\tilde{\varepsilon}_R$/[d$\bar{\eta}$d$\theta$sin($\theta$)]],   and (f)  radiation energy  d$\tilde{\varepsilon_R}$/[d$\theta$sin($\theta$)]. (g) and (h)  correspond to (e) and (f), respectively, for the case including SE. The  sample electron parameters are given in the text, and the laser parameters are the same as in Fig.~\ref{fig1}. }
\label{fig3}
\end{figure}

We proceed discussing the role of SE in shaping RAD. The dynamics and radiation of a sample electron is analysed  in Fig.~\ref{fig3} (an electron at the beam center, $z=L_e/2$ and $x=y=0$, is considered). We follow the mean longitudinal momentum $\bar{p_z}$ and the mean motion direction  $\bar{\theta_e}$ of the electron during interaction with the laser pulse in dependence on the laser phase $\bar{\eta}=(\omega_0 t-k_0z)/(2\pi)$, see Figs.~\ref{fig3}(a) and \ref{fig3}(b), when SE are included or excluded, respectively.   To reproduce SE we repeat the simulation $200$  times for the same initial conditions of the electron. 
The single-electron radiation angle is $\bar{\theta}_e\sim \arctan(m\xi/\bar{p_z})$  ($0\leq\bar{\theta}_e\leq \pi$). The point $\bar{p_z}=0$ corresponds to the electron  reflection, when the emission angle sharply changes from the forward into the backward direction.

In the case without SE, the single-electron radiation angle is well defined at each moment during interaction. 
In contrast, in the case with SE one observes spreading of the average longitudinal momenta and of the emission angles for a large range of laser phases $\bar{\eta}$, see Figs.~\ref{fig3}(a) and \ref{fig3}(b), which stems from the probabilistic character of the photon-emission process. In each possible trajectory shown in Figs.~\ref{fig3}(a) and \ref{fig3}(b), the electron emits photons at different moments and photons of different energies. Consequently, the electron has a different energy and emission angle at the same laser phase $\bar{\eta}$ (same $\xi$) for different trajectories.

The dynamics of the photon emission described above is illustrated in Figs.~\ref{fig3}(c) and \ref{fig3}(d) for two sample trajectories, having a small and a large  energy after the reflection in Fig.~\ref{fig3}(a). Although both of the trajectories begin with the same initial condition, in the first case the number of photon emissions before the reflection ($\bar{\eta}\lesssim 7.5$) is larger. In the second case the electron ``straggles" (does not emit large number of photons) most of the time before the reflection, only emitting a large photon near the reflection $\bar{\eta} \approx 5.5$ \cite{Suppl_material}. Because of the latter, the parameter $\chi$ after the reflection is larger for the first trajectory with respect to the second, yielding significantly more radiative loss for the first trajectory than for the second one, and correspondingly, smaller energy (larger emission angle)  for the first trajectory than for the second one after the reflection \cite{Suppl_material}.  Thus, different probabilistic dynamics of the photon emission, i.e. SE, induces spreading of the photon emission angle at each laser phase.

The angle-resolved radiation intensity and radiation energy  are shown in Figs.~\ref{fig3}(e) and \ref{fig3}(f), respectively, for the case without SE.  In each cycle strongest radiation arises near the peaks of the cycles at a certain angle. The latter are different for each cycle due to the laser pulse shape. Between adjacent radiation peaks, there is a gap in the emission angle corresponding to the weak-field part of the laser cycle. Therefore, the RAD  reveals the laser-cycle structure when SE are neglected. 

The RAD with SE for a single electron is shown in Figs.~\ref{fig3}(g) and \ref{fig3}(h). Since with SE the radiation angle in each laser cycle has a very broad spread shown in  Figs.~\ref{fig3}(b) and \ref{fig3}(g), the gaps in emission angle between adjacent radiation peaks of each cycle are filled out. Consequently, the radiation intensity in this case  shows a single gamma-radiation peak corresponding to the peak of the laser pulse. Note that the discussed qualitative features of the RAD for a sample electron do not depend on the initial position of the electron in the electron beam. The variation of the initial position introduces only a slight modification of quantitative character \cite{Suppl_material}.

 The electron straggling effect in our setup is estimated in Figs.~\ref{fig2}(c) and \ref{fig2}(d) for the emitted photon number integrated over the whole $2\pi$-region of the azimuthal angle. The straggling effect is observed as an enhancement of the emission of high-energy photons when the SE are included, see Fig.~\ref{fig2}(d) for the emitted photon number above 50 MeV. However,   the  electron straggling effect in the considered regime is insignificant and not relevant for the observation of SE.

 \begin{figure}
\includegraphics[width=8cm]{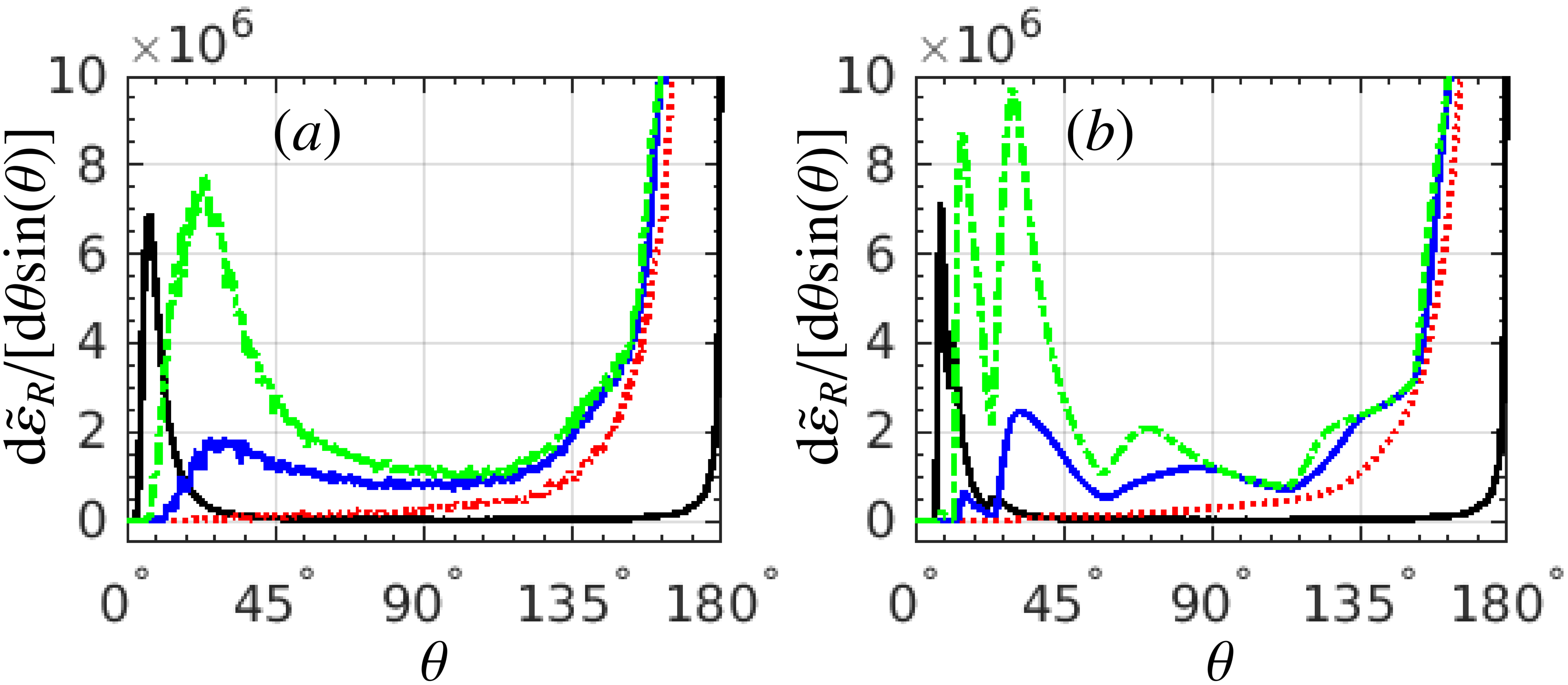}\\
\caption{(Color online) RAD d$\tilde{\varepsilon}_R$/[d$\theta$sin($\theta$)]:  (a) with SE, and (b) without SE, respectively. The laser focal radius $w_0$ equals:  (red, dotted) $\lambda_0$, (blue, solid) 2$\lambda_0$ and (green, dash-dotted)   3$\lambda_0$. The black-solid curves show the case of a 6-cycle plane-wave laser pulse, scaled by a factor of $10^{-2}$. All other laser and electron parameters are the same as in Fig.~\ref{fig1}.}
\label{fig4}
\end{figure} 
\begin{figure}
\includegraphics[width=8cm]{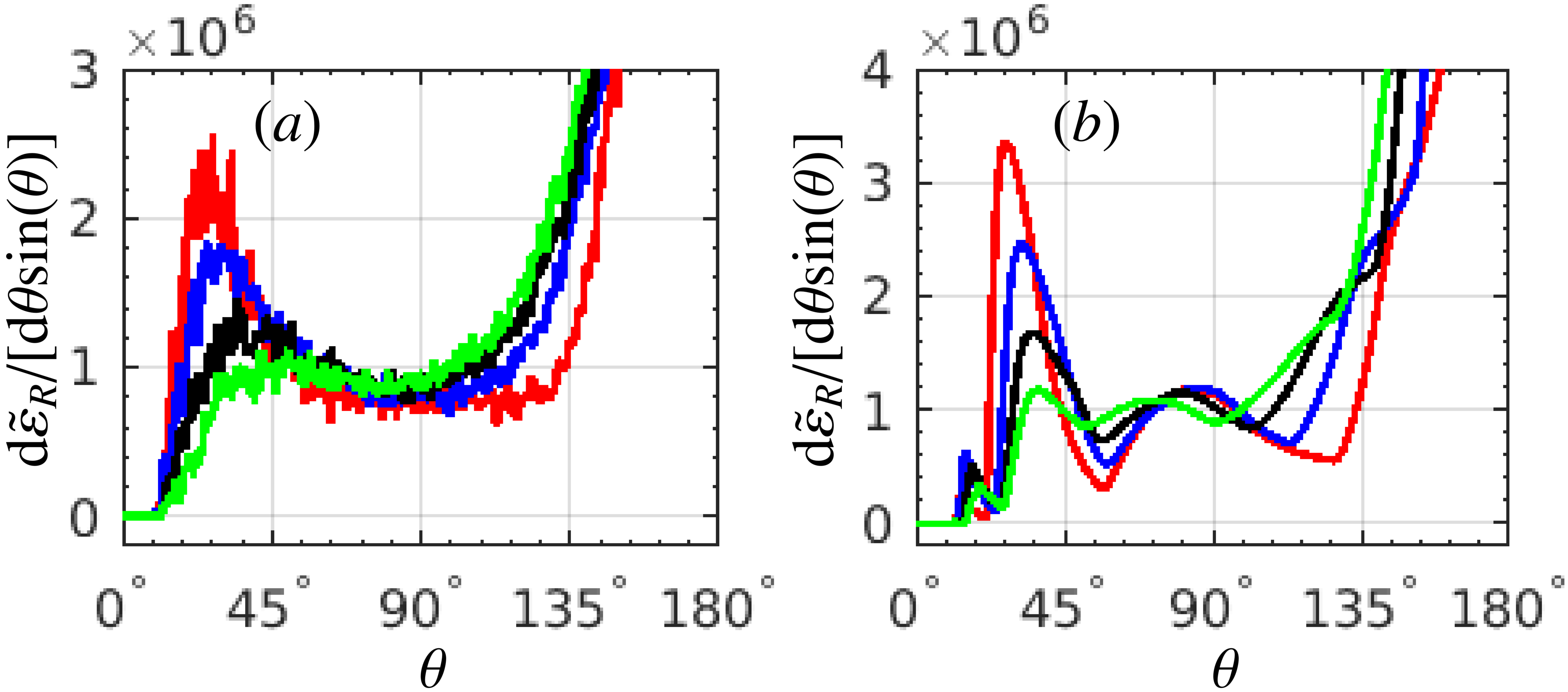}\\
\caption{(Color online) RAD  d$\tilde{\varepsilon}_R$/[d$\theta$sin($\theta$)]:  (a) with SE, and (b) without SE, respectively. The laser-pulse length $L_L$ is: (red) 4$\lambda_0$, (blue) 6$\lambda_0$, (black) 8$\lambda_0$ and (green)  10$\lambda_0$. The electron-beam length $L_e$ is chosen to be equal to the corresponding laser-pulse length.  All other laser and electron parameters are the same as in Fig.~\ref{fig1}.}
\label{fig5}
\end{figure}

We investigate the optimization of the laser (focal radius, pulse duration) and electron (energy) parameters for the best identification of SE.
The influence of the laser focusing effect on RAD is studied in Fig.~\ref{fig4}. The case of  $w_0=2\lambda_0$ is optimal, being the same as in Figs.\ref{fig1} and \ref{fig2}, when the SE are exhibited by a broad smooth peak of the gamma radiation in backward direction. For $w_0=\lambda_0=w_e$, the near-reflection radiation is rather weak, and there is no laser-cycle structure in the ``without SE" case. For $w_0=3\lambda_0$, the main (highest) gamma-photon peaks in the region of $11^{\circ}\lesssim \theta\lesssim 45^{\circ}$  are much stronger apparently than those for $w_0=2\lambda_0$, since all electrons are much closer to the laser-intensity center, which also weakens the shift of the main peaks in polar angle when the SE are excluded. Consequently, higher angular resolution is required to distinguish the ``without SE" case from the one with SE. In the extreme case of a plane-wave laser pulse the structure of multiple gamma-photon peaks in the ``without SE" case cannot be observed any more. Therefore, a tightly focused laser pulse with a focal radius of roughly 2 times of the electron-beam radius applies well for the observation of SE.

The dependence of RAD on the laser-pulse length is shown in Fig.~\ref{fig5}. As the laser-pulse length increases from $4\lambda_0$ to $10\lambda_0$, the main (highest) radiation peak  in the region of $11^{\circ}\lesssim \theta\lesssim 45^{\circ}$  gradually decreases, and the plateaus between the peaks gradually increase. In principle, for a longer laser pulse there are more peaks corresponding to the laser cycles in the ``without SE" case, which is helpful to distinguish SE.  However, for a longer laser pulse, e.g., $L_L=10\lambda_0$, the laser-field gradient is smaller than for shorter laser pulses, consequently, the ratio between the peaks and the plateaus becomes smaller, and the visibility of the signatures becomes worse. Then, a short laser pulse of $L_L\approx 6\lambda_0$ is optimal:  it has a more-cycle structure than for ultrashort pulses, and the visibility of the laser-cycle structure is better than longer pulses.
Finally, we note that higher electron energies are more favourable for SE signature detection as far as $\gamma\sim\xi/2$ is maintained \cite{Suppl_material}.

Concluding, we have revealed   signatures of the stochastic nature of gamma-photon emission during nonlinear Compton scattering of a superstrong short focused laser pulse by a counterpropagating electron beam in the quantum radiation-dominated regime. The signatures are manifested in the qualitative features of the  angular distribution of the radiation in the near-backward direction with respect to the initial electron motion, which arises when  the  electron energy is at the reflection condition. When the stochasticity effects are included, a single broad  gamma-photon peak is formed in the near-reflection radiation angular distribution, while several gamma-photon peaks would arise if there were no stochasticity in the photon emission. The signatures are enhanced with  tightly focused and short laser pulses. They are robust with respect to variation of the laser and electron parameters and can be observed in  near-future laser facilities, such as ELI, HiPER, and XCELS.

\bibliography{strong_fields_bibliography}

\end{document}